\documentclass[prb,twocolumn]{revtex4}
\usepackage[dvips]{graphicx}

\begin{document}

\bibliographystyle{apsrev}

\title{Spatio--temporal dynamics of quantum--well excitons}

\author{Hui Zhao}
\author{B.~Dal Don}
\author{S.~Moehl}
\author{H.~Kalt}

\affiliation{Institut f\"{u}r Angewandte Physik, Universit\"{a}t Karlsruhe, D-76128 Karlsruhe, Germany}

\author{K.~Ohkawa}
\author{D.~Hommel}

\affiliation{Institut f\"{u}r Fesk\"{o}rperphysik, Universit\"{a}t
Bremen, D-28334 Bremen, Germany}

\begin{abstract}
We investigate the lateral transport of excitons in ZnSe quantum
wells by using time--resolved micro-photoluminescence enhanced by
the introduction of a solid immersion lens. The spatial and
temporal resolutions are 200~nm and 5~ps, respectively. Strong
deviation from classical diffusion is observed up to 400~ps. This
feature is attributed to the hot--exciton effects, consistent with
previous experiments under cw excitation. The coupled
transport--relaxation process of hot excitons is modelled by Monte
Carlo simulation. We prove that two basic assumptions typically
accepted in photoluminescence investigations on excitonic
transport, namely (i) the classical diffusion model as well as
(ii) the equivalence between the temporal and spatial evolution of
the exciton population and of the measured photoluminescence, are
not valid for low--temperature experiments.
\end{abstract}

\maketitle

\section{Introduction}

Real--space transport of excitons in semiconductor is an important
aspect of excitonic dynamics and plays an essential role in many
opto--electronic applications. In a quantum--well (QW) structure,
the excitons are confined in the QW plane by the potential
barriers. The two--dimensional lateral transport of excitons in a
number of semiconductor QW structures has been extensively studied
during the past two decades by several optical techniques, e.g.,
transient grating, pump--probe and photoluminescence (PL). In
transient--grating experiments, a spatially periodic distribution
of carrier density is generated by the interference of two laser
pulses with different angles of incidence. The decay of this
periodic distribution is then detected by the diffraction of a
third pulse, and the diffusivity can be deduced from this
decay.\cite{b307346,apl511259,b467528,b473582,b476827,b5116651,b5411046}
In pump--probe measurements, an intense laser pulse is used to
generate excitons, which distribution is then detected by
measuring the absorption of a delayed and much weaker probe beam.
Scanning the probe beam with respect to the pump beam, one can
obtain information on excitonic
transport.\cite{b322407,b385788,b391862,b4613461,b602101,apl7461}

Different from these nonlinear techniques, experiments based on PL
can be performed in the low--density regime, where the
interactions between carriers are negligible. This makes the
interpretation of the experimental results and the deduction of
the physics on exciton--environment coupling more straightforward.
Two types of PL--based methods have been developed during the past
decade. In time--of--flight experiments, the sample surface is
covered by masks. Small holes are etched on this cover to transmit
the excitation laser beam and the luminescence. The decay of the
PL from the hole after a pulsed excitation can be attributed to
both the radiative decay and the transport of excitons out of the
hole, i.e., out of the detection window. Since the radiative decay
can be measured independently by using samples without mask,
information on transport can be deduced by modelling the PL decay
by, normally, diffusion
equation.\cite{apl531973,b3910901,b423220,b451240,jjap325586,ssc88677,b4914523}
The other method is PL imaging, in which the spatial profiles of
the luminescence are measured directly e.g. by a
microscope.\cite{jap852866,physb273963,b626924,apl74741,jap81536,apl74850,jap864697,apl81346}
Extensive information about the carrier diffusivity, mobility and
diffusion length has been extracted based on these two methods.

In these optical investigations based on PL, excitons are excited
by laser photons with an excess energy $E_{\mathrm{excess}}$ of
typically several tens to several hundreds of
meV.\cite{apl531973,b3910901,b423220,b451240,jjap325586,ssc88677,
b4914523,jap852866,physb273963,b626924,apl74741,jap81536,apl74850,jap864697,apl81346}
Despite of the possibility that the transport can happen when the
excitons are still hot, i.e. with high kinetic energy, the
hot--exciton effects are typically neglected in this kind of
studies. For the measurements performed on room
temperature,\cite{jap81536,apl74850,jap864697,apl81346}
hot--exciton effects are less pronounced due to fast relaxation.
However, for low--temperature PL experiments with a rather high
$E_{\mathrm{excess}}$,\cite{apl531973,b3910901,b423220,b451240,jjap325586,ssc88677,
b4914523,jap852866,physb273963,b626924,apl74741} the relaxation
can take several hundreds of picoseconds.\cite{b571390} If hot
excitons play a role in the transport process, two basic
assumptions of these PL experiments are questionable. The first
problem is the validity of the diffusion equation in describing
the transport process. During the relaxation, the kinetic energy
and the group velocity of the hot excitons are decreasing. Thus,
the transport during relaxation cannot be described as diffusion
with a constant diffusivity. Secondly, the measured temporal and
spatial evolution of the PL is {\it not} equivalent to that of the
exciton population. Due to the negligible photon momentum, the PL
can only be used to monitor the cold excitons which have small
enough momentum to be able to couple to a photon (see e.g.
Ref.~\onlinecite{l693393}). The hot excitons are invisible in
these techniques. For this reason, the spatio--temporal evolution
of the total exciton population including hot excitons can be
quite different from that directly deduced from these experiments.

In this paper, we show that these two problems have to be
accounted for in typical PL investigations on the transport of QW
excitons. By performing temporally resolved micro-PL and Monte
Carlo simulation, we investigate the excitonic transport in ZnSe
QWs. Strong deviation from classical diffusion on a time scale of
several hundreds of picoseconds is observed in the
low--temperature experiments and is attributed to a persistent
hot--exciton population. From the simulation, we obtain the
spatio--temporal evolution of the total exciton population and
find a pronounced difference from that directly deduced from the
PL.

\section{Experiment}

The experimental setup is a time--resolved micro--PL ($\mu$--TRPL)
enhanced by introducing a solid immersion lens (SIL). The spatial
and temporal resolutions are 200~nm and 5~ps, respectively.
Figure~1 shows schematically the configuration. The laser system
including an Argon--ion laser and a Ti:Sapphire laser with double
frequency by BBO crystal generates tunable blue laser pulses of
150~fs with a repetition rate of 76~MHz. The pulse shaper reduces
its spectral width to 0.2~meV. The beam is then spatially expanded
to fit the diameter of the objective (magnification 20$\times$,
numerical aperture $\mathrm{NA}=0.4$), and focused on the sample
surface. The luminescence is collected by the same objective, i.e.
confocal configuration, and focused on the image plane of the
microscope. A set of pinholes with different sizes is installed in
this plane, achieving local detection. By moving the pinhole in
the image plane, one can scan the detection spot with respect to
the excitation spot in a well--defined way. The signal
transmitting the pinhole is then spectrally resolved by a
spectrometer, temporally resolved by a streak camera and recorded
by a CCD. Additionally, a shiftable set of mirror and lens is
installed in front of the spectrometer, reflecting and focusing
the light to another CCD camera connected with a monitor. This
configuration achieves a direct imaging of the sample surface on
the monitor, thus ensures fine alignments of the laser beam, the
objective and the pinhole.
\begin{figure}
 \includegraphics[width=8cm]{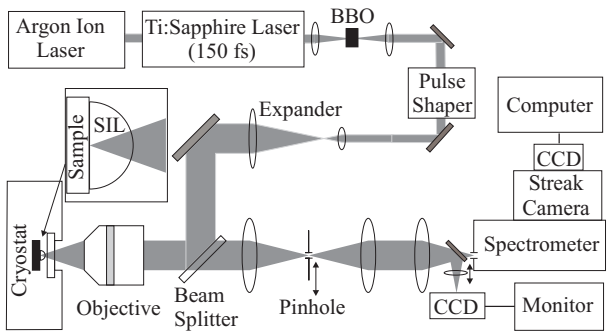}
 \caption{Temporally resolved micro--photoluminescence enhanced by introducing a solid immersion lens (SIL).
 }
\end{figure}

The high spatial resolution is achieved by using a hemispherical
SIL made of Zr$\mathrm{O}_{2}$ with refractive index of
2.16.\cite{apl741791} The SIL is adhesively fixed to the sample
surface. The diameter of the SIL is chosen to be 1~mm, which is
large enough for giving a sufficient field of view, but still
small enough to be stuck on the sample surface adhesively even in
vertical configuration. The SIL enhances not only the spatial
resolution by about two times due to the increased refractive
index of the media outside the sample, but also the collection
efficiency by more than five times due to a larger collection
angle.\cite{jap1} These enhancements are of crucial importance for
the current investigation. All measurements are performed at a
sample temperature of 10~K.

Two ZnSe multiple--QW samples with different periods as well as
different barrier materials are studied. Since similar transport
properties are observed in both samples, we focus here on the
study of a 10--period ZnSe/ZnMgSSe multiple QW with 8~nm well
width and 11~nm barrier width grown by molecular--beam epitaxy.
Figure~2 shows time--integrated (lower panel) and
\begin{figure}
 \includegraphics[width=8cm]{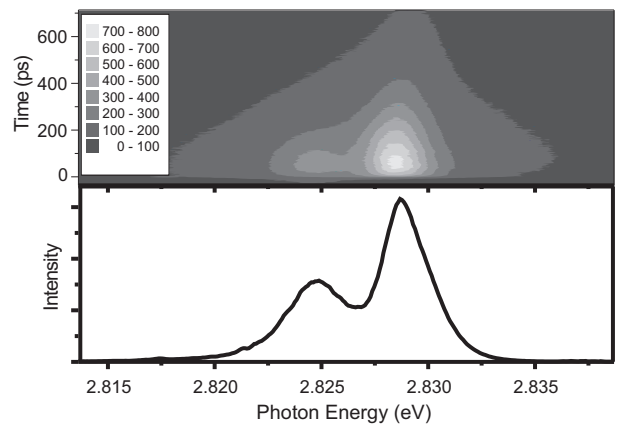}
 \caption{Time--integrated (lower panel) and
time--resolved (upper panel) photoluminescence spectra.
 }
\end{figure}
time--resolved (upper panel) PL spectra of this sample excited by
continuous--wave (cw) laser or pulsed laser, respectively. In both
cases, the energy of the excitation photon is chosen to be
2.9013~eV. The heave--hole exciton resonance is observed at
2.8287~eV with a linewidth of 2.9~meV. The rise time and decay
time of this peak is measured as 65~ps and 205~ps, respectively.
The lower--energy peak is attributed to charged excitons and is
not examined here.

The $\mu$--TRPL spectra are measured by moving the pinhole in the
image plane, thus scanning the detection spot (450~nm in diameter,
defined by the pinhole size) with respect to the fixed excitation
spot. From these spectra, we obtain the temporal evolution of the
spatial profile of heavy--hole exciton luminescence. The detection
window used in this extraction is a small spectral region around
the excitonic resonance.
\begin{figure}
 \includegraphics[width=7cm]{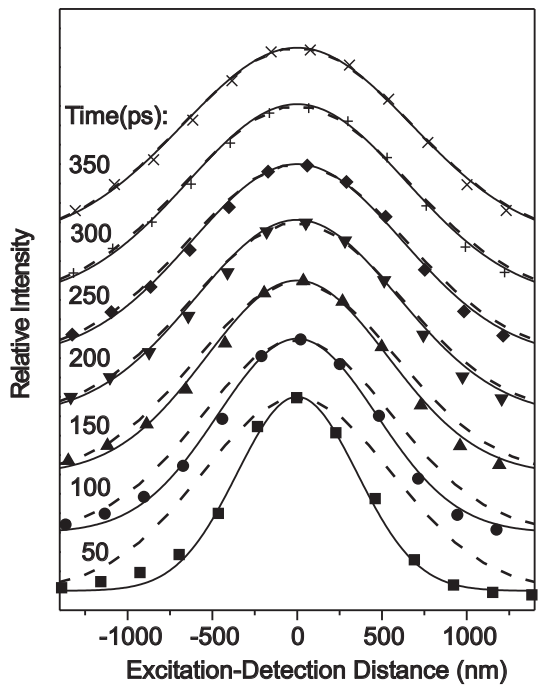}
 \caption{Temporal evolution of the photoluminescence (PL) profile and
 of the exciton population profile. Points with different shapes: The PL profiles at
 different times after the
 excitation measured by scanning the pinhole. Solid (dashed) lines: The corresponding
 PL (total exciton population) profiles obtained by the Monte Carlo simulation.
 }
\end{figure}
Figure~3 shows several PL profiles measured at a set of times
after a pulsed--laser excitation. The photon energy of the laser
is 2.9013~eV. The excitation power is less than 1~$\mu$W, ensuring
the condition of a low exciton density. The expansion of the PL
profile which originates from the real--space transport of
excitons is clearly observed.

\section{Monte Carlo Simulation}

Monte Carlo (MC) simulations\cite{bookmc} have been successfully
applied in investigations on carrier dynamics in semiconductor
QWs.\cite{l59716, b372933, b492177, b5116860, b537322, b5813081,
l831391, b601953} In the present study we use the MC method to
simulate the laser excitation, excitonic transport, relaxation,
and recombination processes. The model used for the simulation is
based on the solid understanding of the hot--exciton formation and
relaxation processes in ZnSe QW. After a laser excitation with a
suitable photon energy, an electron--hole pair is excited in the
continuum states. Generally, the electron--hole pair may (i)
rapidly form a hot exciton with large center--of--mass wavevector,
followed by hot--exciton relaxation, or (ii) dissociate into
individual carriers that relax to their band minima separately. In
GaAs semiconductor structures, both theoretical and experimental
studies have revealed a fast exciton--formation process following
the excitation if the excess energy of the excitation is large
enough for a LO--phonon emission of the excited electron--hole
pair.\cite{b427434,l713878,b5516049,b5813403,b622045,b65035303} In
polar II--VI QWs, the LO--phonon--assisted exciton formation is as
fast as sub--picosecond due to the strong Fr\"ohlich coupling and
thus dominates over the dissociation process. This has been
verified by the experimental observations of hot--exciton
luminescence in these structures.\cite{l67128,b439354,b436843} Our
previous investigations on hot--exciton luminescence in ZnSe QWs
by temporally resolved\cite{b571390} or spatially
resolved\cite{l89097401} LO--phonon--sideband (PSB) spectroscopy
as well as the observation of a pronounced LO--phonon cascade in
PL excitation spectrum\cite{jcg184795} also confirm the dominant
role of this fast exciton--formation process. Thus, in the
simulation, we neglect the dissociation process and assume that
all of the excited electron--hole pairs form excitons by
LO--phonon emission.

In the case that the hot exciton is formed with a center--of--mass
kinetic energy $E_{\mathrm{k}}$ larger than the LO--phonon energy
($E_{\mathrm{LO}}$~=~31.8~meV in ZnSe QW), it relaxes rapidly
toward its band minimum by LO--phonon emissions on a time scale of
100~fs, until $E_{\mathrm{k}}<E_{\mathrm{LO}}$.\cite{jcg184795}
Then the rest of the relaxation is achieved by the much slower
acoustic--phonon emission, and continues over several hundreds of
picoseconds.\cite{b571390} Since the LO--phonon assisted formation
and relaxation are much faster than the rest of the relaxation
process, as a good approximation we can neglect the transport
during the former processes. So the physical process in our
experiment and thus in the model used for the simulation can be
simplified as follows: The laser pulse generates hot excitons with
$E_{\mathrm{k}}=E_{\mathrm{excess}}-nE_{\mathrm{LO}}$. In our
experiment $E_{\mathrm{excess}}$~=~72.6~meV, so $n$~=~2 and
$E_{\mathrm{k}}$~=~9~meV. The initial spatial distribution of the
excitons is defined by the profile of the laser spot. After the
generation, the hot excitons travel in real space and relax in
energy space simultaneously until it recombines at
$E_{\mathrm{k}}\approx 0$. The radiative recombination induces the
PL.

The scattering processes included in our simulation are
acoustic--phonon scattering and interface--roughness scattering.
For the acoustic--phonon scattering, we calculate the scattering
rates at temperature of 10~K by using the model proposed by
Takagahara.\cite{b316552} The differential scattering rates as
functions of scattering angle $\theta$ and $E_{\mathrm{k}}$ are
shown in Figs.~4a and 4b for emission and absorption processes,
respectively. The scattering by acoustic--phonon emission prefers
backward scattering, i.e., the maximum scattering rate is found
for $\theta = \pi$. On the contrary, the acoustic--phonon
absorption prefers forward scattering. Also, at low temperature
the scattering rate of emission is much larger than that of
absorption.
\begin{figure}
 \includegraphics[width=7cm]{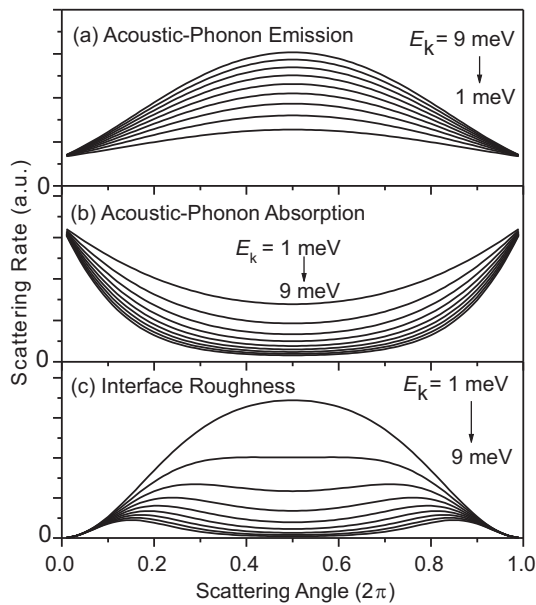}
 \caption{Differential scattering rates of acoustic--phonon
 emission (a), acoustic--phonon absorption (b), and
 interface--roughness scattering (c) as function of scattering
 angle for several exciton kinetic energies $E_{\mathrm{k}}$ from
 1 to 9~meV.
 }
\end{figure}

In a semiconductor QW sample, the well width is not uniform due to
the imperfection of each interface between well and barrier. Such
a fluctuation introduces potential variations which influence the
exciton dynamics, especially the lateral transport process. The
structure of the fluctuation is characterized by a height
$\Delta$, which is normally assumed to be one or two
monolayers,\cite{b537322} and a lateral correlation length
$\lambda$. Since we have no detailed information about the
interface quality of the sample, in our simulation we take the
$\Delta$ to be one monolayer (0.283~nm) and use the $\lambda$ as a
free parameter to fit the experiments. Another choice would be to
leave both of them as free parameters for the fitting. But this
will make the fitting more unreliable due to the one more free
parameter. Since the purpose of the simulation is to describe the
general transport behavior of the excitons, rather than to extract
the information of the interface structure of the sample, we
choose the first option. Thus, due to the uncertainty of the
$\Delta$ used in the calculation, the fitted $\lambda$ shouldn't
be used as a reliable number describing the interface structure.
The interface--roughness scattering rates are calculated by using
the method by Basu et~al,\cite{b441844} as shown in Fig.~4c as
functions of scattering angle and $E_{\mathrm{k}}$. For these
curves, the $\lambda$ is set to be 3.7~nm, which is obtained by
fitting the experiments. The directional property of the
interface--roughness scattering depends strongly on the exciton
energy. For hot excitons the forward scattering is preferred,
while the excitons with smaller energy are more likely to be
scattered backwardly. This feature originates from the fact that
the angular dependence of the differential scattering rate,
\begin{equation}
  (1-\mathrm{cos}\theta) \cdot \mathrm{exp}[-(\lambda
  k)^{2}\mathrm{sin}^{2}\frac{\theta}{2}],
\end{equation}
with $k$ represents the exciton wavevector, is composed of a
forward factor and a backward factor. Thus the $\lambda k$
determines the relative weight of these two factors.

The exciton--exciton scattering is neglected since the experiment
is performed with low exciton density. The optical--phonon
scattering is not included in our simulation since the exciton
energy (9~meV) is too low for an optical--phonon emission and the
optical--phonon absorption is negligible at the sample temperature
of 10~K. For the recombination process, we assume that the exciton
can recombine if its kinetic energy is smaller than the spectral
linewidth (2.9~meV), and the recombination rate $\Gamma$ is
independent of the energy within this recombination
window.\cite{l592337} In the simulation, $\Gamma$ is used as a
second free parameter to fit the experiments. Although the
investigated samples are multiple QWs based on the polar material
ZnSe, the influence of the polariton formation on the transport
process is not included in the simulation. This is based on the
fact that similar transport properties were observed in cw
experiments from both single--QW and multiple--QW
samples.\cite{apl801391} Also in the present time--resolved
experiments, we observe similar transport behaviors from two
multiple--QW samples with different barrier materials as well as
different periods of QW.

At the beginning of the simulation, an ensemble of excitons is
generated by the program. The position of each exciton is defined
by random numbers such that the overall spatial distribution of
the excitons coincides with the profile of the laser spot used in
the experiment. The kinetic energy of the excitons is 9~meV
according to the experiment condition, and the direction of the
velocity of each exciton is selected randomly within the QW plane.
Since the width of the laser pulse (150~fs) is much shorter than
the temporal resolution of our study, the excitons are assumed to
be generated simultaneously. After generation, the exciton travels
in QW plane according to its velocity until it is scattered. The
length of this 'free flight' duration is selected by a random
number generated according to the total scattering rate of all
scattering processes considered. Then one of the scattering
process is chosen to happen, according to the relative scattering
rate among these processes. According to the type of the
scattering process selected, we can determine the state of the
exciton after this scattering event. This state is used as the
initial state of the next 'free flight'. The above procedure is
repeated many times until the radiative recombination happens
which terminates the existence of this exciton. In the simulation,
an ensemble of 30 million excitons is simulated by this method.

\section{Discussion}

As we have discussed, the hot--exciton effects are typically
neglected in the PL investigations on excitonic transport in QW.
In the absence of hot--exciton effects, the exciton population
$n(\vec{r},t)$ obeys the two--dimensional diffusion equation
\begin{equation}
\frac{\partial n(\vec r,t)}{\partial t}=D\nabla^{2}n(\vec
r,t)-\frac{n(\vec r,t)}{\tau}
\end{equation}
with a constant diffusivity $D$.\cite{b385788} The last term
describes the exciton recombination with a lifetime $\tau$. Also,
the measured PL $I(\vec r,t)$ is equivalent with the exciton
population, i.e. $ I(\vec r,t)\propto n(\vec r,t)$, since all of
the excitons are in the recombination window. Based on these
assumptions, the measured temporal evolution of the full--width at
half--maxima (FWHM) of the PL profile obeys\cite{b385788}
\begin{equation}
\mathrm{FWHM^{2}}(t)=\mathrm{FWHM^{2}}(t=0)+16\mathrm{ln}2Dt.
\end{equation}
That is, the square of the FWHM increases linearly with time with
a slope being proportional to $D$.

To compare our experiments with the above formula, we fit the
measured PL profiles in Fig.~3 by Gaussian functions (not shown in
this figure) to get the FWHM's. The obtained temporal evolution of
the $\mathrm{FWHM^{2}}$ is shown in Fig.~5 as the squares. We find
a strong deviation from a linear function, namely a sub--linear
rise of the $\mathrm{FWHM^{2}}$ in the time--range up to 400~ps.
This result shows that the excitonic transport cannot be assumed
as a classical diffusion. It implies the importance of the
hot--exciton effects.
\begin{figure}
 \includegraphics[width=8cm]{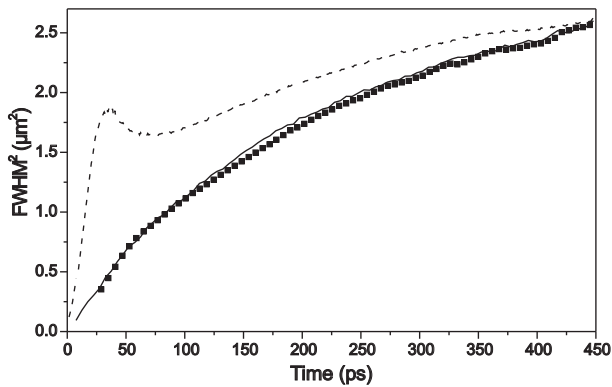}
 \caption{Temporal evolution of the $\mathrm{FWHM}^{2}$ for Gaussian fits to measured photoluminescence profiles (squares),
 the Monte Carlo simulated photoluminescence profiles (solid line), and the Monte Carlo simulated total exciton population (dashed line).
 }
\end{figure}

Beside the nonlinear expansion observe here, we have other
evidences for the important role the hot exciton plays. From a
temporally resolved PSB experiment we have directly observed that
the hot excitons remain non--thermal on a hundred--picosecond time
scale in ZnSe QWs.\cite{b571390} This time scale is long enough
that significant transport can take place. We have also found from
a spatially resolved cw PSB measurements that a non--thermal
distribution of hot excitons can still be observed after a
transport of 1.3~$\mu$m.\cite{apl812794} From a cw measurement on
the zero--phonon line of the PL, we have revealed the importance
of the excitonic kinetic energy on the transport
process.\cite{apl801391} These results together with the nonlinear
expansion observed here by $\mu$--TRPL strongly suggest that the
hot excitons, rather than the relaxed excitons, dominate the
transport.

We want to discuss now the detailed results of the MC simulation
of the hot--exciton transport. As discussed in the previous
section, two free parameters, the correlation length of the
interface roughness $\lambda$ and recombination rate $\Gamma$, are
used in the simulation to fit the experiment. As anticipated,
these two parameters are relatively independent. With other
conditions fixed, the $\lambda$ affects the spatial distribution
of the PL, but has little influence on the temporal evolution. The
opposite is true for the $\Gamma$. With $\Gamma =
5.7\times10^{9}$~/s, the PL decay is fitted satisfactorily (not
shown here). The best fit of the spatial distribution is obtained
when $\lambda = 3.7$~nm, as shown in Fig.~3. Please note that the
solid lines in this figure are not the just mentioned Gaussian
fits to the experimental data, but are obtained by MC simulation
of the excitonic dynamics. Figure~5 provides a more quantitative
comparison of the experiment (squares) and the simulation (solid
line) in terms of FWHM. It is remarkable that a perfect agreement
is achieved by adjusting only one parameter, $\lambda$. This fact
also confirms the validity of our model.
\begin{figure}
 \includegraphics[width=8cm]{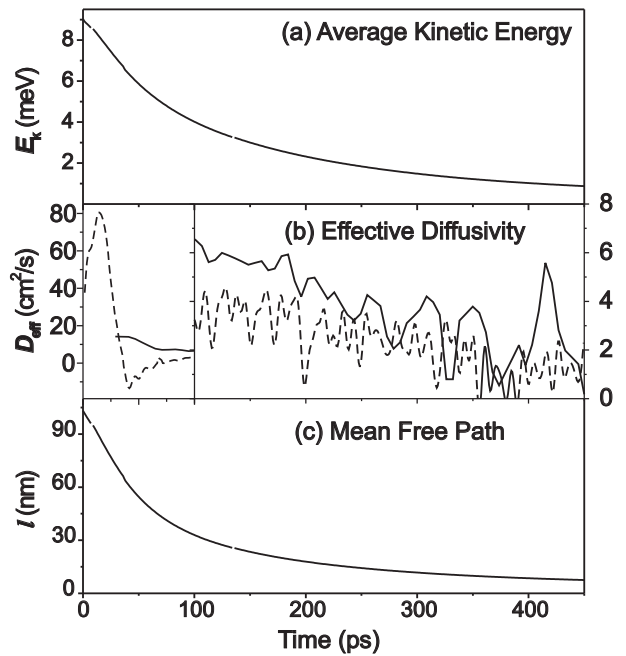}
 \caption{Average kinetic energy (a), effective diffusivity (b),
 and mean free path (c) of the excitons as functions of time. The
 effective diffusivity is obtained by modelling the $I(\vec
r,t)$ (solid line) or $n(\vec r,t)$ (dashed line) with the
modified diffusion model, respectively. Note that different
vertical scales are used for different time--ranges in (b).
 }
\end{figure}

Beside its ability to reproduce the directly observable
experimental results, the simulation also provides us deeper
insight into the underlying carrier dynamics. From the simulation,
one can extract many aspects of the process which are not
accessible in experiments. Figure~6a shows the temporal evolution
of the average kinetic energy of excitons obtained from the
simulation. The energy relaxation from the initial value (9~meV)
to the thermal energy is clearly observed. We find the excitons
remain hot up to about 400~ps. This is consistent with the
sub--linear increase of the $\mathrm{FWHM}^{2}$ up to 400~ps
observed in Fig.~5, and confirms again that the excitonic
transport on a time scale of several hundreds of picoseconds
cannot be described as a classical diffusion with a constant $D$.
After 400~ps, the relaxed excitons show the features of diffusive
transport. On this regime, interface roughness can localize the
relaxed excitons and can result in a sub--diffusive
transport.\cite{b608975} Since the PL decay time is measured as
205~ps (Fig.~2), only a small fraction of the exciton ensemble can
reach this regime. So these effects have minor influence on the
excitonic transport studied here under nonresonant excitation.

If what we proved above were the only problem existing in previous
PL investigations on excitonic transport, one could modify the
diffusion model to fix this problem by simply substituting the
constant diffusivity $D$ by an effective diffusivity varying with
time, $D_{\mathrm{eff}}(t)$.\cite{b4613461} One could even extract
$D_{\mathrm{eff}}(t)$ from the measured PL expansion by applying
Eq.~(3) to intervals which are so small that the energy variation
is negligible. However, a second problem arises from the
hot--exciton effects, as we mentioned in the introduction and will
demonstrate in the following, that such a modified model is still
invalid in describing the PL expansion.

The problem is, that the measured PL $I(\vec r,t)$ is not
equivalent to the exciton population $n(\vec r,t)$ due to the
invisibility of hot excitons in these techniques. Since a large
portion of the exciton ensemble populates high kinetic energy
states, i.e. dark states, during the relaxation, this is quite
obvious. Nevertheless, this problem is typically neglected.
\cite{apl531973,b3910901,b423220,b451240,jjap325586,ssc88677,
b4914523,jap852866,physb273963,b626924,apl74741} From our
simulation, we find that during the relaxation the $n(\vec r,t)$
is significantly different from the measured $I(\vec r,t)$, as
shown in Fig.~3. In other words, the spatial distributions of
excitons shown as the dashed lines in Fig.~3 can result in a
narrower PL profile as shown by the solid lines. In Fig.~5 this
difference is revealed in a quantitative way. The
$\mathrm{FWHM}^{2}$ of the $n(\vec r,t)$ (dashed line) is
significantly larger than that of $I(\vec r,t)$ (solid line)
within about 400~ps. This can be understood by the fact that the
hot excitons with high velocity can travel out of the excitation
spot quite fast, but they are not visible in the spectrum.

In Fig.~5 we find a striking peak around 30~ps. This peak implies
a breathing of the spatial distribution of the hot excitons. That
is, in average the hot excitons initially travel out of the
excitation spot, then move back for a while until they go outward
again. Such an oscillation is in striking contrast to the model of
classical diffusion, and can be understood according to the
directional property of acoustic--phonon scattering. According to
the differential scattering rates shown in Fig.~4a, the scattering
of acoustic--phonon emission prefers a backward scattering.
Initially, the hot excitons are generated with an outward velocity
in average. During their travelling, interface--roughness
scattering events can happen. Since in this transport regime the
energy of the excitons is still high (9~meV), the direction of the
velocity is only slightly changed due to the forward--scattering
character of interface roughness for hot excitons (Fig.~4c). As
soon as an acoustic--phonon scattering event happens, the
direction of the velocity is most likely reversed. Then the
excitons in average move toward the excitation spot, resulting in
the peak observed in Fig.~5. The next acoustic--phonon scattering
event changes the direction of the velocity again. However, since
the scattering events happen to each exciton in a random manner,
the oscillating feature is averaged out after few scattering
events. So, only one peak is observed in Fig.~5. We are aware of
the fact, that we cannot give a direct experimental proof for the
occurrence of the oscillations found in the MC simulations. But,
the important message is here: when such oscillations of the
hot--exciton population happen, they are not detectable in a
standard PL.

To confirm quantitatively the invalidity of the modified diffusion
model in describing the PL expansion, we compare in Fig.~6b the
curves of $D_{\mathrm{eff}}(t)$ extracted by modelling $I(\vec
r,t)$ (solid line) or $n(\vec r,t)$ (dashed line) with the
modified diffusion model, respectively. We find the curves to be
quite different, especially in the first 100~ps. The
$D_{\mathrm{eff}}(t)$ obtained from $n(\vec r,t)$ increases
rapidly in about 30~ps, then drops to a negative value. This
feature is induced by the breathing of the hot--exciton
population. The difference between these two curves becomes
smaller as the hot excitons relax toward the band minimum. After
about 400~ps the two curves overlap, indicating the excitons are
thermalized. This is consistent with the energy relaxation process
obtained from the simulation, as shown in Fig.~6a.

From the simulation, we also extract the mean free path ($l$) for
each exciton. Figure~6c shows the temporal evolution of the
averaged $l$ of the whole ensemble. Since the rate of the
interface--roughness scattering is larger than that of
acoustic--phonon scattering, the $l$ is governed by the
interface--roughness scattering. During the energy relaxation, the
exciton velocity decreases. Also, the interface--roughness
scattering rate increases with decreasing the exciton energy.
These two factors induce the decrease of $l$ with time. We note
that initially the mean free path is nearly 100~nm, reflecting the
ballistic transport right after the hot--exciton generation.

\section{Conclusion}
In conclusion, we show by $\mu$--TRPL and MC simulation that the
excitonic transport in ZnSe QWs at low temperature is dominated by
hot excitons. Consequently, two basic assumptions in previous PL
investigations on excitonic transport: (i) the description by
classical diffusion as well as (ii) the equivalence between the
total exciton population and the measured PL profiles, are not
valid in low--temperature experiments.

\section*{Acknowledgement}
This work was supported by the Deutsche Forschungsgemeinschaft
(DFG) within grant Ka~761/10--1 and within the DFG--Center for
Functional Nanostructures (CFN).

\end{document}